\documentclass[final,journal,11pt]{IEEEtran}
\usepackage{graphicx}
\usepackage{amssymb}
\usepackage{epstopdf}
\usepackage{amsmath,amsthm}
\usepackage{enumerate}
\usepackage{eufrak}
\usepackage{cite}
\usepackage{mathcomp}
\usepackage{supertabular}
\usepackage{longtable}
\usepackage{stmaryrd}
\usepackage{url}
\usepackage{color}
\usepackage{rotating}
\usepackage{float}
\usepackage{multicol,setspace}

\usepackage{tikz,tkz-graph}
\tikzstyle{vertex}=[circle, draw,inner sep =0pt, minimum size=6pt]
\newcommand{\vertex}{\node[vertex]}

\theoremstyle{definition}

\newtheorem{prop}{Proposition}[section]
\newtheorem{thm}{Theorem}[section]
\newtheorem{cor}{Corollary}[section]
\newtheorem{lem}{Lemma}[section]

\newtheorem{defn}{Definition}[section]
\newtheorem{exa}{Example}[section]

\newtheorem*{conjecture}{Conjecture}

\newcommand{\wtp}{{\rm wt_p}}

\newcommand{\etal}{{\em et al.}}

\newcommand{\ZZ}{{\mathbb Z}}

\newcommand{\C}{{\cal C}}

\newcommand{\vu}{{\mathbf u}}
\newcommand{\vv}{{\mathbf v}}

\newcommand{\floor}[1]{\left\lfloor #1 \right\rfloor}

\title{Maximum Distance Separable Codes for Symbol-Pair Read Channels}
\author{Yeow Meng Chee,~\IEEEmembership{Senior Member, IEEE}, Lijun Ji, Han Mao Kiah, %~\IEEEmembership{Student Member, IEEE}, \\
Chengmin Wang, and Jianxing Yin

\thanks{Research of Y.~M.~Chee, H.~M.~Kiah and C.~Wang
is supported in part by the National Research Foundation of Singapore under Research Grant NRF-CRP2-2007-03.
The research of L. Ji is supported by NSFC under Grant 11222113. 
The research of J. Yin and C. Wang is supported by NSFC under Grant 11271280.}

\thanks{Y.~M.~Chee and H.~M.~Kiah  
are with the Division~of~Mathematical Sciences,
  School~of~Physical~and~Mathematical~Sciences, Nanyang~Technological~University, 21~Nanyang~Link, Singapore~637371, Singapore
   (email: ymchee@ntu.edu.sg; kiah0001@ntu.edu.sg).}
   
\thanks{L.~Ji and J.~Yin are with the Department~of~Mathematics, Suzhou~University, Suzhou 215006, China (email: jilijun@suda.edu.cn; jxyin@suda.edu.cn).}
 
\thanks{C.~Wang is with School~of~Science, Jiangnan~University, Wuxi 214122, China (email:wcm@jiangnan.edu.cn).}
\thanks{This paper was presented in part \cite{Cheeetal:2012} at 2012
IEEE International Symposium on Information Theory.}
 }

\begin{document}

\maketitle

\begin{abstract}
%\boldmath
We study (symbol-pair) codes for symbol-pair read channels introduced
recently by Cassuto and Blaum (2010). A Singleton-type bound on symbol-pair codes
is established and infinite families of optimal symbol-pair codes are constructed.
These codes are maximum distance separable (MDS) in the sense that they meet the
Singleton-type bound. 
In contrast to classical codes, where all known $q$-ary MDS codes have length $O(q)$,
we show that $q$-ary MDS symbol-pair codes can have length $\Omega(q^2)$.
%We also construct equidistant cyclic MDS symbol-pair codes from Mendelsohn designs
In addition, we completely determine the existence of MDS symbol-pair codes for certain parameters.
\end{abstract}


\IEEEpeerreviewmaketitle

\section{Introduction}

Symbol-pair coding theory has recently been introduced by Cassuto and Blaum
\cite{CassutoBlaum:2010,CassutoBlaum:2011} 
to address channels with high write resolution but low read resolution, so that
individual symbols cannot be read off due to physical limitations.
An example of such channels is magnetic-storage, where information may be written
via a high resolution process such as lithography and then read off by a low resolution
technology such as magnetic head.

The theory of symbol-pair codes is at a rudimentary stage. Cassuto and Blaum
\cite{CassutoBlaum:2010,CassutoBlaum:2011} laid out a framework for combating
pair-errors, relating pair-error correction capability to a new metric called
pair-distance. They also provided code constructions
and studied decoding methods. Bounds and asymptotics on the size of optimal
symbol-pair codes are obtained. More recently, Cassuto and Litsyn \cite{CassutoLitsyn:2011}
constructed cyclic symbol-pair codes using algebraic methods, and
showed that there exist symbol-pair codes whose rates are strictly higher, compared
to codes for the Hamming metric with the same relative distance.
Yaakobi \etal{} \cite{Yaakobietal:2012} presented efficient algorithms for decoding of  
cyclic symbol-pair codes. 

This paper continues the investigation of codes for symbol-pair channels.
We establish a Singleton-type bound for symbol-pair codes 
and construct MDS symbol-pair codes (codes meeting this Singleton-type bound).
In particular, we construct $q$-ary MDS symbol-pair codes of length $n$ and pair-distance 
$n-1$ and $n-2$,  where $n$ can be as large as $\Omega(q^2)$. 
In contrast, the lengths of nontrivial classical $q$-ary MDS codes are conjectured
to be $O(q)$. 
%In addition, we provide a new construction for equidistant cyclic MDS symbol-pair codes based on Mendelsohn designs.
In addition, we completely settle the existence of MDS symbol-pair codes of length $n$
with pair-distance $d$, for certain parameters.

The rest of the paper is organized as follows. 
In Section 2 we introduce basic notation and definitions and derive a Singleton-type bound for symbol-pair codes.
In Section 3 we make use of interleaving and graph theoretic concepts to construct MDS symbol-pair codes 
from classical MDS codes.
Unfortunately, such methods are inadequate to determine completely the existence of MDS symbol-pair codes.
In Section 4 we introduce other construction methods and give complete solutions in certain instances.
Technical proofs are deferred to the Appendix. %and certain explicit symbol-pair codes 

%the following set of parameters:
%\begin{enumerate}[(i)]
%\item $2\le d\le 4$ and $d=n$, for all $n$,
%\item $d=n-1$, for $6\le n\le 8$, and,
%\item $d=n-2$, for $7\le n\le 10$.
%\end{enumerate}

%Parts of the paper have been presented in \cite{Cheeetal:2012}.
% $2\leq d \leq 4$ and $d=n$. 
%We also construct $q$-ary MDS symbol-pair codes of length $n$ and pair-distance 
%$n-1$ and $n-2$,  where $n$ can be as large as $\Omega(q^2)$. 
%In contrast, the lengths of nontrivial classical $q$-ary MDS codes are conjectured
%to be $O(q)$.
%

%\pagebreak
\section{Preliminaries}

Throughout this paper, $\Sigma$ is a set of $q$ elements, called {\em symbols}.
For a positive integer $n\ge 2$, $\mathbb{Z}_n$ denotes 
the ring $\mathbb{Z}/n\mathbb{Z}$. 
The coordinates of $\mathbf{u}\in\Sigma^n$ are indexed by elements of ${\mathbb Z}_n$, so that
$\mathbf{u}=(u_0,u_1,\cdots, u_{n-1})$.

A {\em pair-vector} over $\Sigma$ is a vector in $(\Sigma\times\Sigma)^n$.
We emphasize that a vector is a pair-vector through the notation $(\Sigma\times\Sigma)^n$,
in lieu of $(\Sigma^2)^n$.
For any $\mathbf{u}=(u_0,u_1,\cdots,u_{n-1})\in \Sigma^n$, the {\em symbol-pair
read vector} of $\mathbf{u}$ is the pair-vector (over $\Sigma$)
{\small
\begin{equation*}
	\pi(\mathbf{u})=((u_0,u_1),(u_1,u_2),\cdots,(u_{n-2},u_{n-1}),(u_{n-1},u_0)).
\end{equation*}
}
Obviously, each vector $\mathbf{u}\in \Sigma^n$ has a unique
symbol-pair read vector $\pi(\mathbf{u})\in(\Sigma\times\Sigma)^n$. 
However, not all pair-vectors over $\Sigma$
have a corresponding vector in $\Sigma^n$.
%If a pair-vector has a corresponding vector, then
% $u_i'=u_{i+1}$ for each $i\in \mathbb{Z}_{n-1}$ and $u_{n-1}'=u_0$.

%Let $\mathbf{u},\mathbf{v}\in(\Sigma\times\Sigma)^n$.
Let $\vu,\vv\in \Sigma^n$.
The {\em pair-distance} between vectors $\mathbf{u}$ 
and $\mathbf{v}$ is defined as
\begin{align*}
	D_\text{p}(\mathbf{u},\mathbf{v})&:=D_\text{H}(\pi(\vu),\pi(\vv))\\
	&=|\{i\in\mathbb{Z}_n: (u_i,u_{i+1}) \neq (v_i,v_{i+1})\}|, 
\end{align*}
%The pair-distance between two vectors in $\Sigma^n$ is the pair-distance between 
%their corresponding symbol-pair read vectors, and if $\mathbf{u},\mathbf{v}\in\Sigma^n$,
%we write
%$D_\text{p}(\mathbf{u},\mathbf{v})$ to mean $D_\text{p}(\pi(\mathbf{u}),\pi(\mathbf{v}))$.
\noindent where $D_\text{H}$ denotes the usual Hamming distance.
Cassuto and Blaum \cite{CassutoBlaum:2011} proved that 
$(\Sigma^n,D_\text{p})$ is a metric space, and showed the following relationship
between pair-distance and Hamming distance. % $D_\text{H}$.

\vskip 5pt
\begin{prop}[Cassuto and Blaum \cite{CassutoBlaum:2011}]
\label{prop:hammingpair}
For $\mathbf{u},\mathbf{v}\in\Sigma^n$ such that $0<D_\text{H}(\mathbf{u},\mathbf{v})<n$,
we have
\begin{equation*}
D_\text{H}(\mathbf{u},\mathbf{v})+1\leq D_\text{p}(\mathbf{u},\mathbf{v})\leq 
2D_\text{H}(\mathbf{u},\mathbf{v}).
\end{equation*}
In the extreme cases in which $D_\text{H}(\mathbf{u},\mathbf{v})=0$ or $n$, we have
$D_\text{p}(\mathbf{u},\mathbf{v})=D_\text{H}(\mathbf{u},\mathbf{v})$.
\end{prop}
\vskip 5pt

A ({\em $q$-ary}) {\em code of length $n$} is a set ${\cal C}\subseteq\Sigma^n$.
Elements of $\cal C$ are called {\em codewords}.
The code $\cal C$ is said to have {\em pair-distance} $d$ 
%if $D_\text{p}(\mathbf{u},\mathbf{v})\geq d$ for all distinct $\mathbf{u},\mathbf{v}\in{\cal C}$, 
if $d=\min \{D_\text{p}(\vu,\vv): \vu,\vv \in \C, \vu\ne\vv \}$
and we denote such
a code by {\em $(n,d)_q$-symbol-pair code}.
The {\em size} of an $(n,d)_q$-symbol-pair code is the number
of codewords it contains and the size of a symbol-pair code satisfies the following Singleton-type bound. 
%The maximum size of an $(n,d)_q$-symbol-pair code is
%denoted $A_q^\text{p}(n,d)$. An $(n,d)_q$-symbol-pair code 
%having $A_q^\text{p}(n,d)$
%codewords is said to be {\em optimal}. The size of an optimal symbol-pair code
%satisfies the following Singleton-type bound.

\vskip 5pt
\begin{thm}\label{thm:bound}[Singleton Bound]
Let $q\geq 2$ and $2\leq d \leq n$. If $\C$ is an $(n,d)_q$-symbol-pair code, then
%$A_q^\text{p}(n,d)\leq q^{n-d+2}$.
$|\C|\le q^{n-d+2}$.
\end{thm}

\begin{IEEEproof}
Let $\cal C$ be an $(n,d)_q$-symbol-pair code with $q\geq 2$ and $2\leq d\leq n$.
Delete the last $d-2$ coordinates
from all the codewords of $\cal C$. Observe that any $d-2$ consecutive
coordinates contribute at most $d-1$ to the pair-distance.
Since $\cal C$ has pair-distance
$d$, the resulting vectors of length $n-d+2$ remain
distinct after deleting the last $d-2$ coordinates from all
codewords. The maximum number of distinct vectors of length $n-d+2$ over an
alphabet of size $q$ is
$q^{n-d+2}$. 
Hence, $|\C| \leq q^{n-d+2}$.
%Hence, $A_q^\text{p}(n,d)\leq q^{n-d+2}$.
\end{IEEEproof}
\vskip 5pt

We call an $(n,d)_q$-symbol-pair code of size $q^{n-d+2}$ \emph{maximum distance separable} (MDS). 
In this paper, we construct new infinite classes of MDS symbol-pair codes and 
completely determine the existence of MDS symbol-pair codes
for certain parameters.

\section{MDS Symbol-Pair Codes from Classical MDS Codes}\label{sec:mds}

In this section, we give several methods for deriving MDS symbol-pair codes
from classical MDS codes. 

%We also provide direct constructions for MDS symbol-pair codes.
Note that ${\cal C}=\Sigma^n$ is trivially an MDS $(n,2)_q$-symbol-pair code for all
$n\geq 2$ and $q\ge 2$ and so, we consider codes of pair-distance at least three.

\subsection{MDS Symbol-Pair Codes and Classical MDS Codes}

Recall that a classical MDS $(n,d)_q$-code,
is a $q$-ary code of length $n$ 
with Hamming distance $d$ and size $q^{n-d+1}$.
Exploiting the relationship between pair-distance and Hamming distance, we develop 
some general constructions for MDS symbol-pair codes
and determine the existence of all such codes with pair-distance three.

\vskip 5pt
\begin{prop}\label{prop:mds}
An MDS $(n,d)_q$-code with $d<n$ is an MDS $(n,d+1)_q$-symbol-pair code.
\end{prop}

\begin{IEEEproof}
Let $\C$ be an MDS $(n,d)_q$-code of size $q^{n-d+1}$. 
By Proposition \ref{prop:hammingpair}, $\C$ has pair-distance at least $d+1$.
Therefore $\C$ meets the Singleton bound of Theorem \ref{thm:bound}.
\end{IEEEproof}
\vskip 5pt

The following corollary follows immediately from classcial MDS codes, mainly, Reed-Solomon codes and their extensions (see \cite{Hedayatetal:1999}).

\vskip 5pt
\begin{cor}\label{cor:rs}
\hfill
\begin{enumerate}[(i)]
\item There exists an MDS $(n,n-1)_{q}$-symbol-pair code for all $q=2^m$, $m\geq 1$ and $n\le q+2$.
\item There exists an MDS $(n,5)_{q}$-symbol-pair code for all $q=2^m$, $m\geq 1$ and $n\le q+2$.
\item There exists an MDS $(n,d)_{q}$-symbol-pair code whenever $q$ is
a prime power, $4\leq d\leq n$ and $n\le q+1$.
\item There exists an MDS $(n,3)_q$-symbol-pair code for all $n\geq 2$, $q\geq 2$.
\end{enumerate}
\end{cor}

\noindent In particular, Corollary \ref{cor:rs}(iv) %implies $q(n,3)=2$.
settles completely the existence of MDS $(n,3)_q$-symbol-pair codes.

\subsection{MDS Symbol-Pair Codes from Interleaving Classical MDS Codes}

We use the interleaving 
method of Cassuto and Blaum \cite{CassutoBlaum:2011} to 
obtain MDS symbol-pair codes. 
Cassuto and Blaum showed that a symbol-pair code with even pair-distance 
can be obtained by interleaving two classical codes of the same length and distance. 

\vskip 5pt
\begin{thm}[{Cassuto and Blaum\cite{CassutoBlaum:2011}}]\label{thm:interleaving}
If there exist an $(n,d)_q$-code of size $M_1$ and an $(n,d)_q$-code of size $M_2$, 
then there exists a $(2n,2d)_q$-symbol-pair code of size $M_1M_2$.
\end{thm}
\vskip 5pt

Interleaving classical MDS codes yield the following corollary.

\vskip 5pt
\begin{cor}\label{cor:rsinterleave}
\hfill
\begin{enumerate}[(i)]
\item There exists an MDS $(2n,2n-4)_{q}$-symbol-pair code for all $q=2^m$, $m\geq 1$ and $n\le q+2$.
\item There exists an MDS $(2n,8)_{q}$-symbol-pair code for all $q=2^m$, $m\geq 1$ and $n\le q+2$.
\item There exists an MDS $(2n,2d)_{q}$-symbol-pair code whenever $q$ is
a prime power, $3\leq d\le n-1$ and $n\le q+1$.
\item There exists an MDS $(2n,4)_q$-symbol-pair code for all $n\geq 2$, $q\geq 2$.
\item There exists an MDS $(2n,2n)_q$-symbol-pair code for all $n\geq 2$, $q\geq 2$.
\end{enumerate}
\end{cor}

Corollary \ref{cor:rsinterleave} (iv) and (v) settle the existence of MDS $(n,4)_q$-symbol-pair codes
and MDS $(n,n)_q$-symbol-pair codes for even $n$. 
In Section \ref{sec:dir} we exhibit that such MDS codes indeed exist for all $n\ge 2$ and $q\ge 2$.

\subsection{MDS Symbol-Pair Codes from Extending Classical MDS Codes}

MDS symbol-pair codes obtained by interleaving necessarily have even length and distance. 
Furthermore, the length of symbol-pair codes obtained is only a factor of two longer than that of
the input classical codes.
In this subsection, we use graph theoretical concepts to extend 
classical MDS codes of length $n$ to MDS symbol-pair codes of length up to $n(n-1)/2$.

We use standard concepts of graph theory presented by Bondy and Murty \cite[chap. 1--3]{BondyMurty:2008}.
Namely, a graph is a pair $G=(V, E)$, where $V$ is a set of {\em vertices}
and $E$ is a set of unordered pairs of $V$, called {\em edges}.	
The {\em order} of $G$ is $|V|$, the number of vertices, while 
the {\em size} of $G$ is $|E|$, the number of edges.

A {\em trail} of length $k$ in $G$ is a list of vertices $v_0,v_1, \ldots, v_k$ such that
$\{v_i,v_{i+1}\}\in E$ for $0\leq i\leq k-1$, and $\{v_i,v_{i+1}\}\not=\{v_j,v_{j+1}\}$ for
$0\leq i<j\leq k-1$. The trail is said to be {\em closed} if $v_0=v_k$. A closed trail
$v_0,v_1,\ldots,v_k$ is a {\em cycle} if $v_i\not=v_j$ for $0\leq i<j\leq k-1$.
%A trail $T=v_0,v_1,\ldots,v_k$ specifies a subgraph $G(T)=(V',E')$, where
%$V'=\{v_i:0\leq i\leq k\}$ and $E'=\{\{v_i,v_{i+1}: 0\leq i\leq k-1\}$.
The length of a shortest cycle in a graph is called its {\em girth}.

On the other hand, a trail that transverses all edges in $G$ is called an {\em eulerian trail}.
If $G$ admits a closed eulerian trail, then $G$ is said to be {\em eulerian}. 
Equipped with the concepts of girth and eulerian trails, we introduce the next construction. 
  
%and assume readers' familiarity.
%A graph is a pair $G=(V, E)$, where $V$ is a set of {\em vertices}
%and $E$ is a set of 2-subsets of $V$, called {\em edges}.	
%The {\em order} of $G$ is $|V|$, the number of vertices.
%The {\em complete graph} $K_n$ is the graph of order $n$ in which every 2-subset of vertices
%is an edge.
%A {\em subgraph} of $G$ is a graph $G'=(V',E')$ such that $V'\subseteq V$ and $E'\subseteq E$.
%A {\em trail} of length $k$ in $G$ is a list of vertices $v_0,v_1, \ldots, v_k$ such that
%$\{v_i,v_{i+1}\}\in E$ for $0\leq i\leq k-1$, and $\{v_i,v_{i+1}\}\not=\{v_j,v_{j+1}\}$ for
%$0\leq i<j\leq k-1$. The trail is said to be {\em closed} if $v_0=v_k$. A closed trail
%$v_0,v_1,\ldots,v_k$ is a {\em cycle} if $v_i\not=v_j$ for $0\leq i<j\leq k-1$.
%A trail $T=v_0,v_1,\ldots,v_k$ specifies a subgraph $G(T)=(V',E')$, where
%$V'=\{v_i:0\leq i\leq k\}$ and $E'=\{\{v_i,v_{i+1}: 0\leq i\leq k-1\}$.
%The length of a shortest cycle in $G(T)$ is called the {\em girth}
%of the trail $T$.

\vskip 5pt
\begin{prop}\label{prop:extmds}
Suppose there exists an MDS $(n,d)_q$-code and there exists an eulerian graph of 
order $n$, size $m$ and girth at least $n-d+1$.
Then there exists an MDS $(m,m-n+d+1)_q$-symbol-pair code.
\end{prop}

\begin{IEEEproof}
Let $G$ be an eulerian graph of order $n$,
size $m$ and girth at least $n-d+1$, where $V(G)=\ZZ_n$.
Consider a closed eulerian trail
$T=x_0e_1x_1e_2x_2\cdots e_mx_m$, where $x_m=x_0$, $x_i\in V(G)$, and
$e_i\in E(G)$, for $1\leq i\leq m$.
Let $\C$ be an MDS $(n,d)_q$-code and
consider the $q$-ary code of length $m$,
\begin{equation*}
\C'=\{ (u_{x_0},u_{x_1},\ldots,u_{x_{m-1}}):\mathbf{u}\in\C\}.
\end{equation*}
We claim that $\C'$ has pair-distance at least $m-n+d+1$. 
Indeed,
pick any $\mathbf{u},\mathbf{v}\in \C$. Since $D_{\rm H}(\mathbf{u},\mathbf{v})\geq d$,  
we have $|\{x\in V(G): u_x=v_x\}|\leq n-d$.   
It follows that
{\small
\begin{equation*}
|\{i: (u_{x_i},u_{x_{i+1}})=(v_{x_i},v_{x_{i+1}}), 0\leq i\leq m-1\}| \leq n-d-1,
\end{equation*}
}
since on the contrary
there would exist at least $n-d$ edges 
$\{y_1,z_1\},\{y_2,z_2\},\ldots,\{y_{n-d},z_{n-d}\}$ in $E(G)$
such that $u_{y_j}=v_{y_j}$ and $u_{z_j}=v_{z_j}$ for all $ 1\leq j\leq n-d$. 
But since the number of vertices $x\in V(G)$ such that $u_x=v_x$ is at most $n-d$, 
these $n-d$ edges must induce a subgraph (of order $n-d$) that contains a cycle of length
at most $n-d$.
This contradicts our assumption
that $G$ has girth at least $n-d+1$.

Consequently, $D_{\rm p}(\mathbf{u},\mathbf{v})\geq m-n+d+1$. 
Finally, observe that $|\C'|=|\C|=q^{n-d+1}$, and hence
$\C'$ is an MDS symbol-pair code by Theorem \ref{thm:bound}.
\end{IEEEproof}
\vskip 5pt

\begin{exa}\label{exa:extend}
Consider the complete graph $K_5$ of order five, whose vertex set of is $\ZZ_5$. 
Hence its edge set comprises all ten unordered pairs of $\ZZ_5$.
The graph $K_5$ is eulerian as it admits the closed eulerian trail, 01234024130.
Trivially, the girth of $K_5$ is three.
 
Hence, given an MDS $(5,3)_q$-code $\C$ and 
since $K_5$ satisfies the conditions of Proposition \ref{prop:extmds},
we have an MDS $(10,9)_q$-symbol-pair code.

More concretely, an MDS $(10,9)_q$-symbol-pair code is given by
\begin{equation*}
\C'=\{ (u_0,u_1,u_2,u_3,u_4,u_0,u_2,u_4,u_1,u_3):\mathbf{u}\in\C\}.
\end{equation*}

Observe that when $q=4$, an MDS $(10,9)_4$-symbol-pair code cannot be obtained 
via Corollary \ref{cor:rs} or Corollary \ref{cor:rsinterleave}.
\end{exa}

\vskip 5pt

To apply Proposition \ref{prop:extmds}, 
we need eulerian graphs of specified order, size, and girth.
However,
little is known about how many edges an eulerian graph with a given
number of vertices and given girth can have.
Nov\'ak \cite{Novak:1971,Novak:1974} proved tight upper bounds on the number of edges
in an eulerian graph of girth four. Below,
we establish the following results on the size of an eulerian graph of order $n$ (of girth three),
and those of girth four.

\vskip 5pt
\begin{prop}
\label{evengraphs}
Let $n\geq 3$ and $M=n\lfloor (n-1)/2\rfloor$.
Then there exists an eulerian graph of order $n$ and size $m$, for
$n\leq m\leq M$, except when $m\in\{M-1,M-2\}$.
\end{prop}
\vskip 5pt

Define
\begin{equation*}
M(n)=\begin{cases}
2\lfloor n^2/8\rfloor,&\text{if $n$ even} \\
2\lfloor (n-1)^2/8\rfloor +1,&\text{if $n$ odd}.
\end{cases}
\end{equation*}

\vskip 5pt
\begin{prop}
\label{girth4}
Let $n\geq 6$.
Then there exists an eulerian graph of order $n$, size $m$, and girth at least four,
for all $m\equiv n\bmod{2}$, $n\leq m\leq M(n)$, except when $m=M(n)-2$ and $n\ge 8$.
\end{prop}
\vskip 5pt

%\begin{IEEEproof}[Idea of proofs for Proposition \ref{evengraphs} and \ref{girth4}]
%An extremal eulerian graph $G$ with girth at least three (or four) of size $M$ (respectively, $M(n)$) can be constructed directly.
%Hamiliton cycles and cycles of appropriate lengths are then removed from $G$ to obtain eulerian graphs of the required sizes.
%It remains to ensure that the subgraph remains connected after the cycles are removed.
%Details are deferred to the full version of this paper.
%\end{IEEEproof}

For $m\not\equiv n\bmod{2}$, we have the following.

\begin{prop}
\label{girth4opp}
\hfill
\begin{enumerate}[(i)]
\item For even $n\geq 10$, there exists an eulerian graph of order $n$, girth at least four, and 
size $m\in\{M(n-2)-1,M(n-2)+1\}$.
\item For odd $n\geq 9$, there exists an eulerian graph of order $n$, girth at least four, and 
size $m\in\{M(n)-1,M(n)-3\}$.
\end{enumerate}
\end{prop}
\vskip 5pt

We remark that Nov\'ak \cite{Novak:1971,Novak:1974} established the existence of eulerian graph of order $n$ and 
girth at least four with size exactly $M(n)$. In contrast, Proposition \ref{girth4} and Proposition \ref{girth4opp} 
provide an eulerian graph of order $n$ and girth at least four
for a spectrum of sizes.
Proofs for Proposition \ref{evengraphs}, Proposition \ref{girth4} and Proposition \ref{girth4opp} are deferred to Appendix \ref{sec:graph}.

\begin{cor}\label{cor:extend1}
Let $q$ be a prime power, $q\geq 4$. Then
there exists an MDS $(n,n-1)_q$-symbol-pair code whenever
\begin{enumerate}[(i)]
\item $2\leq n\leq (q^2-1)/2-3$ or $n=(q^2-1)/2$, for $q$ odd;
\item $2\leq n\leq q(q+2)/2-3$ or $n=q(q+2)/2$, for $q$ even.
\end{enumerate}
\end{cor}

\begin{IEEEproof}
%Follows from Corollary \ref{cor:rs}, Proposition \ref{prop:extmds} and Proposition \ref{evengraphs}.
Apply Proposition \ref{prop:extmds} and Proposition \ref{evengraphs} to classical MDS codes. %Reed-Solomon codes and their extensions.
\end{IEEEproof}
\vskip 5pt

\vskip 5pt
\begin{cor}\label{cor:extend2}
Let $q$ be a prime power, $q\geq 5$. Then
there exists an MDS $(n,n-2)_q$-symbol-pair code whenever
\begin{enumerate}[(i)]
\item $2\leq n\leq M(q)+1$, or $M(q)+1\leq n\leq M(q+1)$ and $n$ even and $n\ne M(q+1)-2$, for $q$ odd;
\item $2\leq n\leq q^2/4+1$, $n\not=q^2/4-1$, for $q$ even.
\end{enumerate}
\end{cor}

\begin{IEEEproof}
%Follows from Corollary \ref{cor:rs}, Proposition \ref{prop:extmds},
%Proposition \ref{girth4} and Proposition \ref{girth4opp}.
Apply Proposition \ref{prop:extmds},
Proposition \ref{girth4} and Proposition \ref{girth4opp}
to classical MDS codes.
%Reed-Solomon codes and their extensions.
\end{IEEEproof}
\vskip 5pt

These results show that in contrast to classical $q$-ary MDS codes of length $n$, 
where it is conjectured that $n\leq q+2$, we can have $q$-ary MDS symbol-pair codes
of length $n$ with $n=\Omega(q^2)$.

\section{Construction of MDS Symbol-Pair Codes with Specific Lengths and Pair-Distances 
and the Existence of MDS Symbol-Pair Codes}\label{sec:dir}

Observe that while Section \ref{sec:mds} constructs MDS symbol-pair codes from classical MDS codes,
the latter is usually defined over a finite field, whose size is necessarily a prime power.   
Unfortunately, the set of prime powers has density zero in the set of positive integers.

In contrast, for fixed $n$ and $d$, we conjecture that the set of alphabet sizes where 
an MDS $(n,d)_q$-symbol-pair code exists has density one. 
Specifically, we conjecture the following.

\begin{conjecture}
Fix $2\le d\le n$. There exists a $q_0$ such that an MDS $(n,d)_q$-symbol-pair exists for all $q\ge q_0$.
\end{conjecture}

%We verify the conjecture for the abovementioned parameters.
In this section, we verify the conjecture for the following set of parameters:
\begin{enumerate}[(i)]
\item $2\le d\le 4$ and $d=n$, for all $n$,
\item $d=n-1$, for $6\le n\le 8$, and,
\item $d=n-2$, for $7\le n\le 10$.
\end{enumerate}

To this end, we utilize a recursive method that builds an MDS symbol-pair code over a larger alphabet
using MDS symbol-pair codes defined over smaller alphabets.
This recursive construction is introduced formally in Subsection \ref{sec:recur}.
However, to seed this recursion, the MDS symbol-pair codes given in Section \ref{sec:mds} are insufficient.
Therefore, we need additional MDS $(n,d)_q$-symbol-pair codes, particularly when $q$ is not a prime power.  
These codes are given in Subsection \ref{sec:linear} and Subsection \ref{sec:nonlinear}.

\subsection{$\mathbb{Z}_q$-linear MDS Symbol-Pair Codes}\label{sec:linear}

We provide constructions for MDS $(n,d)_q$-symbol-pair codes for $d\in\{4,5,n\}$
and for certain small values of $n$, $d$ and $q$.
We remark that for even $n$, MDS $(n,4)_q$-symbol-pair codes have been 
constructed in Corollary \ref{cor:rsinterleave}, 
and MDS $(n,n)_q$-symbol-pair codes can be constructed by interleaving classical repetition codes. 
Here, we construct MDS $(n,4)_q$-symbol-pair codes and MDS $(n,n)_q$-symbol-pair codes for all $n$.

Throughout this subsection, we assume $\Sigma=\mathbb{Z}_q$.
Besides being MDS, the codes constructed have {\em $\ZZ_q$-linearity}.

\vskip 5pt
\begin{defn}
A code
$\C\subseteq\Sigma^n$ is said to be {\em $\ZZ_q$-linear} if 
$\mathbf{u}+\mathbf{v}, \lambda \mathbf{u} \in \C$ for all 
$\mathbf{u}, \mathbf{v}\in \C$ and $\lambda\in \ZZ_q$.
\end{defn}
\vskip 5pt
As with classical codes, a $\ZZ_q$-linear code must contain the zero vector ${\bf 0}$. In addition, 
determining the minimum pair-distance of a $\ZZ_q$-linear code is equivalent to determining the 
minimum {\em pair-weight} of a nonzero codeword. 

\vskip 5pt
\begin{defn}
The {\em pair-weight} of $\mathbf{u}\in\Sigma^n$ is $\text{wt}_\text{p}(\mathbf{u})=D_\text{p}(\mathbf{u},\mathbf{0})$.
%\begin{equation*}
%\text{wt}_\text{p}(\mathbf{u})=D_\text{p}(\mathbf{u},\mathbf{0}).
%\end{equation*}
\end{defn}
\vskip 5pt
The proof of the following lemma is similar to the classical case.

\vskip 5pt
\begin{lem}\label{lem:wt}
Let $\C$ be a $\ZZ_q$-linear code. 
Then $\C$ has pair-distance $\min_{\mathbf{u}\in \C\setminus \{{\mathbf{0}}\}} \wtp(\mathbf{u})$.
\end{lem}
\vskip 5pt

In the rest of the subsection, the $\ZZ_q$-linear codes we construct are of size $q^k$.
We describe such a code via a {\em generator matrix in standard form}, 
that is, a $k\times n$ matrix over $\ZZ_q$ of the form,
\begin{equation*}
G=(I_k | X),
\end{equation*} 
\noindent so that each codeword is given by $\vu G$, where $\vu\in\ZZ_q^k$.

\vskip 5pt 
\begin{prop}\label{prop:d=4}
Let $n\ge 4$ and $q\ge 2$. Let $\C$ be a $\ZZ_q$-linear code with generator matrix,
\begin{equation*}
G=\left(
\begin{array}{cccccc}
1 & 0 & \cdots & 0 & 1 & 1\\
0 & 1 & \cdots & 0 & 2 & 1\\
\vdots & \vdots & \ddots & \vdots & \vdots & \vdots \\
0 & 0 & \cdots & 1 & n-2 & 1\\
\end{array}
\right).
\end{equation*}
%Let $n\ge 4$ and define $f$ and $g$  as follows:
%\begin{align*}
%f: \mathbb{Z}_q^{n-2} &\longrightarrow \mathbb{Z}_q \\
%\mathbf{u} &\longmapsto \sum_{i=0}^{n-3} (i+1)u_i, \\
%g: \mathbb{Z}_q^{n-2} &\longrightarrow \mathbb{Z}_q \\
%\mathbf{u} &\longmapsto \sum_{i=0}^{n-3} u_i.
%\end{align*}
%Let $\C=\{(u_0,u_1,\ldots,u_{n-3}, f(\mathbf{u}),g(\mathbf{u})): \mathbf{u}\in \ZZ_q^{n-2}\}$. 
Then $\C$ is a $\ZZ_q$-linear MDS $(n,4)_q$-symbol-pair code. 
\end{prop}

\begin{IEEEproof}
It is readily verified that $\C$ has size $q^{n-2}$. 
Hence, by Lemma \ref{lem:wt}, it suffices to show that for all 
$\mathbf{u}\in\ZZ_q^{n-2}\setminus \{\mathbf{0}\}$, 
%\begin{equation*}
%\text{wt}_\text{p}((u_0,u_1,\ldots,u_{n-3}, f(\mathbf{u}),g(\mathbf{u})))\geq 4.
%\end{equation*}
\begin{equation*}
\text{wt}_\text{p}(\vu G)\geq 4.
\end{equation*}

Write {\small $\tilde {\mathbf{u}}=\left(u_0,u_1,\ldots,u_{n-3}, \sum_{i=0}^{n-3}(i+1)u_i,\sum_{i=0}^{n-3}u_i\right)$ }
and let
{\small
\begin{align*}
\Delta		&=	\{\text{$i: 0\leq i\leq n-3$ and $u_i\not= 0$}\},\\
\Delta_\text{p}	&=\{i: \text{$0\leq i\leq n-4$ or $i=n-1$, and $(u_i,u_{i+1})\not= \mathbf{0}$}\}.
\end{align*}
}
We have the following cases.
\begin{enumerate}[(i)]
\item {\em The case $|\Delta|\ge 3$}: \\
Then $|\Delta_\text{p}|\geq 4$, and so $\text{wt}_\text{p}(\tilde{\mathbf{u}})\geq 4$. 

\item {\em The case $|\Delta|=2$}: \\
If $\Delta\not=\{j,j+1\}$ for all $0\leq  j\leq n-4$, 
then $|\Delta_\text{p}|\geq 4$, and so $\text{wt}_\text{p}(\tilde{\mathbf{u}})\geq 4$.
If $\Delta=\{j,j+1\}$ for some $j$, $0\leq j\leq n-3$,
then either $\tilde\vu_{n-2}$ or $\tilde\vu_{n-1}$ is nonzero. Otherwise, 
\begin{align*}
(j+1)u_j+(j+2)u_{j+1} & =  0,\\
u_j+u_{j+1} & =  0,
\end{align*}
which implies $u_{j+1}= 0$, a contradiction. 
Hence,  $|\Delta_\text{p}|\geq 3$, and since 
$\tilde\vu_{n-2}$ or $\tilde\vu_{n-1}$ is nonzero, $\text{wt}_\text{p}(\tilde{\mathbf{u}})\geq 4$.

\item {\em The case $|\Delta|=1$}: \\
If $u_0\ne 0$, then both $\tilde\vu_{n-2}$ and $\tilde\vu_{n-1}$ are nonzero. 
Hence, $\text{wt}_\text{p}(\tilde{\mathbf{u}})\geq 4$. If $u_j\neq 0$ for some $j$, $1\leq j\leq n-3$, 
then $\tilde\vu_{n-1}$ is nonzero and 
$\{j-1,j,n-2,n-1\}\subseteq\{i: (u_i,u_{i+1})\not= \mathbf{0}\}$ 
and hence, $\text{wt}_\text{p}(\tilde{\mathbf{u}})\geq 4$.
\end{enumerate}
This completes the proof.
\end{IEEEproof}

\vskip 5pt

\begin{prop}\label{prop:d=n}
Let $n\geq 2$ and 
let $\C$ be a $\ZZ_q$-linear code with generator matrix,
{\small
\begin{equation*}
G= \begin{cases}
\left(
\begin{array}{ccccccc}
1 & 0 & 1 & 0 &\cdots & 1 & 0\\
0 & 1 & 0 & 1 &\cdots & 0 & 1
\end{array}
\right),
& \mbox{ if $n$ is even,} \\
\left(
\begin{array}{cccccccc}
1 & 0 & 1 & 0 &\cdots & 1 & 0 & 1\\
0 & 1 & 0 & 1 &\cdots & 0 & 1 & 1
\end{array}
\right),
& \mbox{ otherwise.} \\
\end{cases}
\end{equation*}
}
%let
%\begin{equation*}
%\C=\begin{cases}
%\{(i,j,i,j,\ldots,i,j):(i,j)\in\ZZ_q^2\},&\text{if $n$ is even} \\
%\{(i,j,i,j,\ldots,i,j,i+j):(i,j)\in\ZZ_q^2\},&\text{if $n$ is odd.}
%\end{cases}
%\end{equation*}
Then $\C$ is an MDS $(n,n)_q$-symbol-pair code.
\end{prop}

\begin{IEEEproof}
It is readily verified that $\C$ has size $q^2$.
Hence, by Lemma \ref{lem:wt}, it is also easy to see that the pair-weight 
of all nonzero vectors in $\C$ is $n$.
\end{IEEEproof}

\vskip 5pt

Proposition \ref{prop:d=4} and Proposition \ref{prop:d=n} settle completely the existence
of MDS $(n,4)$- and $(n,n)$-symbol-pair codes respectively. 
When $5\le d\le n-1$, the task is complex and hence, we determine the existence only for a certain set of parameters.

The next two propositions provide an infinite class and some small MDS symbol-pair codes required to seed the recursive method
in Section \ref{sec:recur}.

 \begin{prop}\label{prop:d=5}
Suppose that $q$ is odd prime and $5\leq n\leq 2q+3$.
Let $\C$ be a $\ZZ_q$-linear code with generator matrix,
\begin{equation*}
G=\left(
\begin{array}{cccccccc}
1 & 0 & 0 & \cdots & 0 & 1 & 1 & 1 \\
0 & 1 & 0 & \cdots & 0 & 2 & 1 & -1\\
0 & 0 & 1 & \cdots & 0 & 3 & 1 & 1\\

\vdots & \vdots & \vdots & \ddots & \vdots & \vdots & \vdots & \vdots\\
0 & 0 & 0 & \cdots & 1 & n-3 & 1 & (-1)^{n-4}\\
\end{array}
\right).
\end{equation*}

%Let $\C=\{(u_0,u_1,\ldots,u_{n-4}, f(\mathbf{u}),g(\mathbf{u}),h(\mathbf{u})): \mathbf{u}\in
%\ZZ_q^{n-3}\}$. 
Then $\C$ is an MDS $(n,5)_q$-symbol-pair code.
\end{prop}

%\begin{IEEEproof}
%Similar to the proof for Proposition \ref{prop:d=4}.
%\end{IEEEproof}
%\vskip 5pt

\begin{IEEEproof}
It is readily verified that $\C$ has size $q^{n-3}$. Hence, by
Lemma \ref{lem:wt}, it suffices to show that for all
$\vu\in\ZZ_q^{n-3}\setminus \{{\bf 0}\}$, \[
\wtp(\vu G)\ge 5.\] 

Define
$f$, $g$ and $h$ as follows:
\begin{align*}
f: \mathbb{Z}_q^{n-3} &\longrightarrow \mathbb{Z}_q, &
\mathbf{u} &\longmapsto \sum_{i=0}^{n-4} (i+1)u_i, \\
g: \mathbb{Z}_q^{n-3} &\longrightarrow \mathbb{Z}_q, &
\mathbf{u} &\longmapsto \sum_{i=0}^{n-4} u_i,\\
h: \mathbb{Z}_q^{n-3} &\longrightarrow \mathbb{Z}_q, &
\mathbf{u} &\longmapsto \sum_{i=0}^{n-4} (-1)^iu_i.
\end{align*}
%\begin{align*}
%f: \mathbb{Z}_q^{n-3} &\longrightarrow \mathbb{Z}_q \\
%\mathbf{u} &\longmapsto \sum_{i=0}^{n-4} (i+1)u_i, \\
%g: \mathbb{Z}_q^{n-3} &\longrightarrow \mathbb{Z}_q \\
%\mathbf{u} &\longmapsto \sum_{i=0}^{n-4} u_i.\\
%h: \mathbb{Z}_q^{n-3} &\longrightarrow \mathbb{Z}_q \\
%\mathbf{u} &\longmapsto \sum_{i=0}^{n-4} (-1)^iu_i.
%\end{align*}
Write
$\tilde {\vu}=(u_0,u_1,\ldots,u_{n-4}, f(\vu),g(\vu),h(\vu))$ and
let %$\Delta$ be the support of $\vu$, or
\begin{align*}
\Delta &=\{i: 0\le i\le n-4, u_i\ne 0\},\\
\Delta_\text{p} &=\{i: i\in\ZZ_n, (\tilde \vu_i,\tilde \vu_{i+1}) \ne \bf 0\}
\end{align*}

\noindent We have the following cases.
\begin{enumerate}[(i)]
\item \emph{The case $|\Delta|\ge 4$}:

Then $|\Delta_\text{p}|\ge 5$ and so, $\wtp(\tilde{\vu})\ge 5$.

\item \emph{The case $|\Delta|=3$}:

If $\Delta\ne\{j,j+1,j+2\}$ for all $0\le  j\le n-6$, then $|\Delta_\text p|\ge 5$ and so
$\wtp(\tilde{\vu})\ge 5$. Otherwise, $\Delta=\{j,j+1,j+2\}$ for some
$0\le j\le n-6$. Then either $g(\vu)$ or $h(\vu)$ is nonzero.
Otherwise,
\begin{align*}
u_j+u_{j+1}+u_{j+2} & =  0,\\
u_j-u_{j+1}+u_{j+2} & =  0,
\end{align*}
implies that $2u_{j+1}= 0$. Since $q$ is odd, $u_{j+1}=0$, a
contradiction. Hence, %$|\{i: 0\le i\le n-1, (u_i,u_{i+1})\ne {\bf0}\}|\ge 5$ and then 
$\wtp(\tilde{\vu})\ge 5$.

\item \emph{The case $|\Delta|=2$}:

(1) Suppose that $\Delta=\{i,j\}$ with $j-i>1$.

If $j-i\equiv 1$ (mod 2), then either $g(\vu)$ or $h(\vu)$ is
nonzero. so $\wtp(\tilde{\vu})\ge 5$. Otherwise,
\begin{align*}
u_i+u_j & =  0,\\
u_i-u_j & =  0,
\end{align*}
implies that $2u_i=0$. Since $q$ is odd, $u_i=0$, a contradiction.

If $j-i\equiv 0$ (mod 2), then either $f(\vu)$ or $g(\vu)$ is
nonzero, so $\wtp(\tilde{\vu})\ge 5$. Otherwise,
\begin{align*}
(i+1)u_i+(j+1)u_j & =  0,\\
u_i+u_j           & =  0.
\end{align*}
implies that $(j-i)u_j=0$. Since $j-i\leq n-4\leq 2q-1$ is even and
$q$ is prime, $u_j=0$, a contradiction.

(2) Suppose that $\Delta=\{j,j+1\}$ for some $0\le j\le n-5$.

If $j=0$, then either $f(\vu)$ or $g(\vu)=0$ and hence, $\wtp(\tilde{\vu})\ge 5$.
Otherwise, $j>0$, then either $g(\vu)$ or $h(\vu)=0$ and so, $\wtp(\tilde{\vu})\ge 5$.
%If $j=0$ and $g(\vu)$ is nonzero, then $\wtp(\tilde{\vu})\ge 5$. If
%$j=0$ and $g(\vu)=0$, then it is readily verified that $g(\vu)$ is
%nonzero, hence $\wtp(\tilde{\vu})\ge 5$. Similarly, if $j>0$, then
%either $g(\vu)$ or $h(\vu)$ is nonzero, so $\wtp(\tilde{\vu})\ge 5$.

\item \emph{The case $|\Delta|=1$}:

If $u_0\ne 0$, then both $f(\vu)$ and $g(\vu)$ are nonzero. So,
$\wtp(\tilde{\vu})\ge 5$. Otherwise, $u_j\ne 0$ for some $1\le j\le
n-4$. Then both $g(\vu)$ and $h(\vu)$ are nonzero and hence,
$\wtp(\tilde{\vu})\ge 5$.
\end{enumerate}
This completes the proof. %Since $\C$ has size $q^{n-3}$, $\C$ is MDS.
\end{IEEEproof}

\vskip 5pt

\begin{prop}\label{prop:linear}
There exist $\ZZ_q$-linear MDS $(n,d)_q$-symbol-pair codes for the following set of parameters,
{
\begin{enumerate}[(i)]
\item $q=2$, $(n,d)\in\{(6,5), (7,6), (7,5), (8,6), (9,7)$, $(10,8)\}$,
\item $q=3$, $(n,d)\in\{(7,6),(8,7),(9,7),(10,8)\}$,
\item $q=5$, $(n,d)=(9,7)$.
\end{enumerate}
}
\end{prop}

\begin{IEEEproof}
Generator matrices for the respective codes are given in Table \ref{tab:linear}.
\begin{table*}
\centering
\caption{Generator Matrices for $\ZZ_q$-linear MDS Symbol-Pair Codes}
\label{tab:linear}
\renewcommand{\arraystretch}{1.3}
\setlength{\tabcolsep}{8pt}
\begin{tabular}{cccp{5.5cm} cccp{5.5cm}}
\hline
$q$ & $n$ & $d$ & Generator matrix for a $\ZZ_q$-linear MDS $(n,d)_q$-symbol-pair code &
$q$ & $n$ & $d$ & Generator matrix for a $\ZZ_q$-linear MDS $(n,d)_q$-symbol-pair code \\
\hline
2 & 6 & 5 &
$\left(
\begin{array}{ c c c c c c }
1 & 0 & 0 & 1 & 0 & 1 \\
0 & 1 & 0 & 1 & 1 & 0 \\
0 & 0 & 1 & 1 & 1 & 1 \\
\end{array}
\right)$ 
&

2 & 7 & 6 &
$\left(
\begin{array}{ c c c c c c c }
1 & 0 & 0 & 1 & 0 & 1 & 1 \\
0 & 1 & 0 & 1 & 1 & 1 & 0 \\
0 & 0 & 1 & 0 & 1 & 1 & 1 \\
\end{array}
\right)$ \\

 & 7 & 5 &
$\left(
\begin{array}{ c c c c c c c }
1 & 0 & 0 & 0 & 1 & 0 & 1 \\
0 & 1 & 0 & 0 & 1 & 1 & 1 \\
0 & 0 & 1 & 0 & 1 & 1 & 0 \\
0 & 0 & 0 & 1 & 0 & 1 & 1 \\
\end{array}
\right)$ &

 & 8 & 6 &
$\left(
\begin{array}{ c c c c c c c c }
1 & 0 & 0 & 0 & 1 & 0 & 1 & 0 \\
0 & 1 & 0 & 0 & 0 & 1 & 0 & 1 \\
0 & 0 & 1 & 0 & 1 & 0 & 0 & 1 \\
0 & 0 & 0 & 1 & 0 & 1 & 1 & 1 \\
\end{array}
\right)$ \\

 & 9 & 7 &
$\left(
\begin{array}{ c c c c c c c c c }
1 & 0 & 0 & 0 & 1 & 0 & 1 & 1 & 0 \\
0 & 1 & 0 & 0 & 0 & 1 & 0 & 1 & 1 \\
0 & 0 & 1 & 0 & 1 & 1 & 0 & 1 & 0 \\
0 & 0 & 0 & 1 & 1 & 1 & 1 & 1 & 1 \\
\end{array}
\right)$ &
 & 10 & 8 &
$\left(
\begin{array}{ c c c c c c c c c c }
1 & 0 & 0 & 0 & 1 & 0 & 1 & 0 & 1 &
1 \\
0 & 1 & 0 & 0 & 1 & 1 & 0 & 1 & 1 &
0 \\
0 & 0 & 1 & 0 & 1 & 0 & 0 & 1 & 0 &
1 \\
0 & 0 & 0 & 1 & 1 & 1 & 1 & 1 & 1 &
1 \\
\end{array}
\right)$ \\
\hline
3 & 7 & 6 &
$\left(
\begin{array}{ c c c c c c c }
1 & 0 & 0 & 2 & 2 & 1 & 1 \\
0 & 1 & 0 & 0 & 1 & 0 & 1 \\
0 & 0 & 1 & 0 & 1 & 1 & 2 \\
\end{array}
\right)$ &

3 & 8 & 7 &
$\left(
\begin{array}{ c c c c c c c c }
1 & 0 & 0 & 1 & 1 & 1 & 2 & 0 \\
0 & 1 & 0 & 0 & 1 & 1 & 1 & 2 \\
0 & 0 & 1 & 1 & 1 & 2 & 0 & 1 \\
\end{array}
\right)$ \\

 & 9 & 7 &
$\left(
\begin{array}{ c c c c c c c c c }
1 & 0 & 0 & 0 & 2 & 2 & 1 & 0 & 1 \\
0 & 1 & 0 & 0 & 2 & 0 & 1 & 1 & 1 \\
0 & 0 & 1 & 0 & 2 & 2 & 0 & 0 & 2 \\
0 & 0 & 0 & 1 & 1 & 0 & 2 & 1 & 1 \\
\end{array}
\right)$ &
 & 10 & 8 &
$\left(
\begin{array}{ c c c c c c c c c c }
1 & 0 & 0 & 0 & 1 & 1 & 1 & 2 & 2 &
0 \\
0 & 1 & 0 & 0 & 0 & 1 & 1 & 1 & 1 &
2 \\
0 & 0 & 1 & 0 & 0 & 2 & 0 & 1 & 2 &
2 \\
0 & 0 & 0 & 1 & 1 & 1 & 2 & 2 & 1 &
2 \\
\end{array}
\right)$ \\
\hline
5 & 9 & 7 &
$\left(
\begin{array}{ c c c c c c c c c }
1 & 0 & 0 & 0 & 1 & 0 & 1 & 0 & 1 \\
0 & 1 & 0 & 0 & 0 & 1 & 0 & 1 & 1 \\
0 & 0 & 1 & 0 & 1 & 0 & 2 & 0 & 3 \\
0 & 0 & 0 & 1 & 0 & 1 & 0 & 2 & 3 \\
\end{array}
\right)$ \\
\hline
\end{tabular}
\end{table*}
\end{IEEEproof}

\subsection{A Family of MDS Symbol-Pair Codes via Development}\label{sec:nonlinear}

We construct an MDS $(8,7)_{2p}$-symbol-pair code for all odd primes $p$.
Similar to the concept of generator matrices, we obtain a full set of codewords
by {\em developing} a smaller subset of codewords over some group.
The concept of development is ubiquitous in combinatorial design theory 
(see \cite[chap. VI and VII ]{Bethetal:1999})
and we construct the required MDS symbol-pair codes via this method.

We define the notion of development formally. 
Proofs in this subsection are deferred to Appendix \ref{sec:develop}.

\begin{defn}\label{def:g2dev}
Let $n$ be even and $\Gamma$ be an abelian additive group. 
A {\em $\Gamma^2$-development $(n,n-1)$-symbol-pair code} is 
a set of $q$ codewords in $\Gamma^n$ such that
for distinct codewords $\vu$, $\vv$, the following hold,
\begin{enumerate}[(i)]
%\item $u_i-u_{i+2} \ne v_{i}-v_{i+2}$ for $i\in\ZZ_n$,
\item $(u_i-u_j,u_{i+1}-u_{j+1})\ne (v_i-v_j,v_{i+1}-v_{j+1})$ 
for $i,j\in\ZZ_n$, $i\equiv j\bmod 2$, and,
\item $(u_i-u_{j+1},u_{i+1}-u_{j})\ne (v_i-v_{j+1},v_{i+1}-v_{j})$ 
for $i,j\in\ZZ_n$, $i\not\equiv j\bmod 2$.
\end{enumerate}
\end{defn}
\vskip 5pt

\begin{prop} \label{prop:g2dev}
Let $n$ be even.
Suppose $\C_0$ is a $\Gamma^2$-development $(n,n-1)$-symbol-pair code with $|\Gamma|=q$.

For $\vu\in\C_0$, $\alpha,\alpha' \in \Gamma$, let
%{
%\begin{equation}\label{eq:phi}
%\phi(\vu,\alpha,\alpha')=(u_0+\alpha,u_1+\alpha',u_2+\alpha,u_3+\alpha',\ldots,u_{n-2}+\alpha,u_{n-1}+\alpha')
%\end{equation}
%}
{
\begin{multline}\label{eq:phi}
\phi(\vu,\alpha,\alpha')=(u_0+\alpha,u_1+\alpha',\\ u_2+\alpha,u_3+\alpha',\ldots,u_{n-2}+\alpha,u_{n-1}+\alpha')
\end{multline}
}

Then $\C=\{\phi(\vu,\alpha,\alpha'):\vu\in\C_0,\alpha,\alpha' \in \Gamma\}$ is
an MDS $(n,n-1)_q$-symbol-pair code.
\end{prop}

Therefore, to construct an MDS $(n,n-1)_q$-symbol-pair code, it suffices to construct 
a set of $q$ codewords, instead of a set of $q^2$ codewords.
Hence, for certain values of $n$ and $q$, a computer search is effective to construct MDS symbol-pair codes.
In the instance when $n=8$, we have the following collection of $\Gamma^2$-development
MDS $(8,7)_q$-symbol-pair codes.

\begin{prop}\label{prop:2p}
Let $p$ be prime with $p\ge 5$ and $\Gamma=\ZZ_p\times \ZZ_2$.

Let $\C_0$ consist of the following four codewords,
\begin{align*}
((0,0),(0,0), (0,0),(1,0), (0,0),(1,1), (0,0),(0,1)),\\
((0,0),(0,0), (0,1),(1,1), (2,0),(0,1), (2,1),(2,0)),\\
((0,0),(0,0), (1,0),(0,0), (1,1),(0,0), (0,1),(0,0)),\\
((0,0),(0,0), (1,1),(0,1), (0,1),(2,0), (2,0),(2,1)).
\end{align*}
Let $\C_1$ be the following set of $2p-4$ codewords,
\begin{align*}
((0,0),(0,0), (a,0),(\hat a,1), 	(3a,1),(0,1), (2a,1),(2\hat a,0)),\\
((0,0),(0,0), (a,1),(a,0), 		(0,1),(3a,1), (2a,0),(2a,1)),
\end{align*}
\noindent where $a\in\{2,3,\ldots, p-1\}$ and 
\begin{equation*}
\hat a=\begin{cases}
p-1,	&\mbox{if $a=2$,}\\
a-1, &\mbox{otherwise.}
\end{cases}
\end{equation*}
\noindent Then $\C=\C_0\cup \C_1$ is a $\Gamma^2$-development $(8,7)$-symbol-pair code.
\end{prop}

\vskip 5pt

In addition, when $p=3$, a $\ZZ_6^2$-development $(8,7)$-symbol-pair code is 
given by the following six codewords,
\begin{align*}
( 0 , 0 , 0 , 0 , 0 , 0 , 0 , 0 ),\\
( 0 , 0 , 1 , 1 , 0 , 5 , 1 , 2 ),\\
( 0 , 0 , 2 , 2 , 4 , 5 , 3 , 4 ),\\
( 0 , 0 , 3 , 3 , 0 , 4 , 2 , 5 ),\\
( 0 , 0 , 4 , 4 , 2 , 3 , 5 , 1 ),\\
( 0 , 0 , 5 , 5 , 0 , 1 , 4 , 3 ).
\end{align*}

Therefore, applying Proposition \ref{prop:g2dev} and Proposition \ref{prop:2p}, 
we have the following existence result.

\begin{cor} \label{cor:2p}
There exists an MDS $(8,7)_{2p}$-symbol-pair code for odd primes $p$.
\end{cor}

\subsection{Complete Solution of the Existence of MDS Symbol-Pair Codes for certain Parameters}\label{sec:recur}

We settle completely the existence of MDS symbol-pair codes for certain parameters.%In particular, we show the following.

In particular, define 
\begin{multline*}
q(n,d)=\min \{q_0: \mbox{an MDS $(n,d)_q$-symbol-pair code}\\
\mbox{ exists for all $q\ge q_0$}\},
\end{multline*}
\noindent and we establish the following.
\begin{thm}\label{thm:exist}
The following hold.
\begin{enumerate}[(i)]
\item $q(n,d)=2$ for $2\le d\le 4$ and $n\ge d$, or $d=n$,
\item $q(n,n-1)=2$ for $n\in\{6,7\}$, $q(8,7)=3$ and,
\item $q(n,n-2)=2$ for $7\le n\le 10$.
\end{enumerate}
\end{thm}

Observe that Theorem \ref{thm:exist} (i) follows from
the opening remark in Section \ref{sec:mds}, Corollary \ref{cor:rs}(iv), Proposition \ref{prop:d=4} and Proposition \ref{prop:d=n}.
For Theorem \ref{thm:exist} (ii) and Theorem \ref{thm:exist} (iii), we require the following recursive construction.

\begin{prop}[Product Construction]\label{prop:product}
If there exists an MDS $(n,d)_{q_1}$-symbol-pair code and an MDS $(n,d)_{q_2}$-symbol-pair code,
then there exists an MDS $(n,d)_{q_1q_2}$-symbol-pair code.
\end{prop}

\begin{IEEEproof}
Let $\C_i$ be an MDS $(n,d)_{q_i}$-symbol-pair code over $\Sigma_i$ for $i=1,2$.
For $\vu\in\C_1$ and $\vv\in\C_2$, let $\vu\times \vv=((u_0,v_0),(u_1,v_1),\ldots,(u_{n-1},v_{n-1}))\in (\Sigma_1\times \Sigma_2)^n$.

Consider the code $\C$ over $\Sigma_1\times\Sigma_2$,
\begin{equation*}
\C =\{\vu\times\vv: \vu\in \C_1,\vv\in\C_2\}.
\end{equation*}
It is readily verified that $|\C|=(q_1q_2)^{n-d+2}$ and it remains to verify that the minimum pair-distance is at least $d$.

Indeed for distinct $(\vu\times\vv), (\vu'\times\vv')\in \C$,
\begin{align*}
D_\text p(\vu\times\vv,\vu'\times\vv') 
& \ge \max\{D_\text p(\vu,\vu'),D_\text p(\vv,\vv')\} \ge d.
\end{align*}
\end{IEEEproof}

\begin{IEEEproof}[Proof of Theorem \ref{thm:exist}(ii) and (iii)]
Define
{\footnotesize
\begin{align*}
Q(2) &=\{p: p \mbox{ prime}\},\\
Q(3) &=\{p: p\ge 3 \mbox{ prime}\} \cup \{2p:  p\ge 3 \mbox{ prime}\} \cup \{2^r: r\ge 2\}.
\end{align*}
}
To show that $q(n,d)\le q_0$ ($q_0\in\{2,3\}$), it suffices by Proposition \ref{prop:product}  
to construct MDS $(n,d)_q$-symbol-pair codes for $q\in Q(q_0)$.
The required MDS $(n,d)_q$-symbol-pair codes are constructed in Section \ref{sec:mds},
Subsection \ref{sec:linear} and  Subsection \ref{sec:nonlinear}. 
We summarize the results in Table \ref{tab:mds}. 

Observe that $q(n,d)\ge 2$ trivially. However, when $(n,d)=(8,7)$,
regard an $(8,7)_2$-symbol-pair code as a (classical) $(8,7)_4$-code, 
whose size is at most seven by Plotkin bound.
Hence, an MDS $(8,7)_2$-symbol-pair code whose size is eight cannot exist and so, $q(8,7)\ge 3$.
\end{IEEEproof}

\begin{table}[hbtp]
\centering
\caption{Some MDS Symbol-Pair Codes}
\label{tab:mds}
\renewcommand{\arraystretch}{1.2}
\setlength{\tabcolsep}{5pt}
\begin{tabular} {cccl}
\hline
$n$ & $d$ & $q$ & Authority\\
\hline
6 & 5 & 2 & Proposition \ref{prop:linear}\\
& & $p$ odd prime & Proposition \ref{prop:d=5}\\
\hline
7 & 6 & 2,3 & Proposition \ref{prop:linear}\\
& & $p\ge 5$, odd prime & Corollary \ref{cor:extend1}\\
\hline
8 & 7 & 3 & Proposition \ref{prop:linear}\\
& & $p\ge 5$, odd prime & Corollary \ref{cor:extend1}\\
& & $2p$, $p$ odd prime & Corollary \ref{cor:2p}\\
& & $2^r$, $r\ge 2$ & Corollary \ref{cor:extend1}\\

\hline
7 & 5 & 2 & Proposition \ref{prop:linear}\\
& & $p$, $p$ odd prime & Proposition \ref{prop:d=5}\\

\hline
8 & 6 & 2 & Proposition \ref{prop:linear}\\
& & $p$, $p$ odd prime & Corollary \ref{cor:rsinterleave}\\
\hline

9 & 7 & 2,3,5 & Proposition \ref{prop:linear}\\
& & $p\ge 7$, $p$ odd prime & Corollary \ref{cor:extend2}\\
\hline

10 & 8 & 2,3 & Proposition \ref{prop:linear}\\
& & $p\ge 5$, $p$ odd prime & Corollary \ref{cor:rsinterleave}\\
\hline

\end{tabular}
\end{table}

\section{Conclusion}

In this paper, we established a Singleton-type bound for symbol-pair codes 
and constructed infinite families of optimal symbol-pair codes. All these codes are of the
{\em maximum distance separable} (MDS) type in that they meet the
Singleton-type bound. We also show how classical MDS codes can be extended to
MDS symbol-pair codes using eulerian graphs of specified girth. In contrast with
$q$-ary classical MDS codes, where all known such codes have length $O(q)$, we establish
that $q$-ary MDS
symbol-pair codes can have length $\Omega(q^2)$.
In addition, we gave complete solutions to the existence of MDS symbol-pair codes for certain parameters.

\appendices

\section{Eulerian Graphs of Specified Size and Girth} \label{sec:graph}

We give detailed proofs of Proposition \ref{evengraphs}, Proposition \ref{girth4} and Proposition \ref{girth4opp}. 
In particular, we construct eulerian graphs with girth at least three and four and specified sizes.

A graph $G=(V,E)$ is said to be {\em even} if the degree of each vertex is even. 
Hence, we have the following characterization of eulerian graphs due to Euler.

\begin{thm}\label{thm:euler} (see {\cite[Theorem 3.5]{BondyMurty:2008}})
Let $G$ be a connected graph. Then $G$ is eulerian if and only if $G$ is an even graph.
\end{thm}

Next, we define certain operations on graphs which aid us in constructing even graphs. 
\begin{itemize}
\item Let $G,H$ be graphs defined on the same vertex set $V$. We denote the graph $(V,E(G)\cup E(H))$ by $G\cup H$ and 
the graph $(V,E(G)\setminus E(H))$ by $G\setminus H$. Suppose $G$ and $H$ are even graphs. 
If $G$ and $H$ are edge-disjoint, then $G\cup H$ is even and if in addition, $G\cup H$ is connected, then eulerian by Theorem \ref{thm:euler}.
Similarly, if $E(G)\supset E(H)$, then $G\setminus H$ is even and eulerian (if $G\setminus H$ is connected). 
\item Let $G=(V,E)$ be a graph with vertices $u,v$ and edge $e=\{u,v\}$. 
We {\em subdivide edge $e$} (see \cite[\S 2.3]{BondyMurty:2008}) to obtain the graph 
$(V\cup \{ x\}, E\setminus \{e\} \cup \{\{u,x\},\{v,x\}\})$. 
In other words, we add the vertex $x$ and replace the edge $\{u,v\}$ with the edges $\{u,x\}$ and $\{v,x\}$.
Suppose $G$ is an eulerian graph with order $n$, size $m$ and girth $g$.
Then subdividing any edge of $G$, we obtain an eulerian graph with order $n+1$, size $m+1$ and girth at least $g$.
\end{itemize}
With these operations, we prove the stated propositions.
\vskip 5pt

\begin{IEEEproof}[Proof of Proposition \ref{evengraphs}]

The proposition is readily verified for $n\in\{3,4\}$.
When $n\ge 5$, let $k=\floor{(n-1)/2}$ 
and we prove the proposition by induction. 
We first construct eulerian graphs of small sizes
and then inductively add edge-disjoint Hamilton cycles to obtain eulerian graphs of the desired sizes.

Define the following collection of $k$ edge-disjoint Hamilton cycles in $K_n$.

\begin{itemize}
\item When $n=2k+1$, let $V=\ZZ_{2k}\cup \{\infty\}$. For $0\le i\le k-1$, the Hamilton cycle $\Phi_i$ is given by
\[ \Phi_i=(\infty,i,i-1,i+1,\ldots, i-k+1,i+k-1,i-k).\]
\item When $n=2k+2$, let $V=\ZZ_{2k+1}\cup \{\infty\}$. For $0\le i\le k-1$, the Hamilton cycle $\Phi_i$ is given by
\[ \Phi_i=(\infty,i,i-1,i+1,\ldots,i-k,i+k).\]
\end{itemize}

For $3\le m\le 2n-3$, there exist two Hamilton cycles $\Phi_{m_1}$, $\Phi_{m_2}$ and a subgraph $H_m$ such that the following holds,
\begin{enumerate}[(i)]
\item $H_m$ is a subgraph of $\Phi_{m_1}\cup \Phi_{m_2}$,
\item $H_m$ is even with size $m$ and when $m\ge n$, $H_m$ is connected and hence, eulerian.
\end{enumerate}
\noindent We give explicit constructions of $\Phi_{m_1}$, $\Phi_{m_2}$, $H_m$ in Table \ref{tab:decomp1}.

\begin{table*}[htbp]
\centering
\caption{Eulerian graphs of small size with order $n$, girth at least three}
\label{tab:decomp1}
\renewcommand{\arraystretch}{1.3}
\setlength{\tabcolsep}{3pt}
\begin{tabular}{cccl}
\multicolumn{4}{l}{$n=2k+1$, $V=\ZZ_{2k}\cup\{\infty\}$}\\
\hline
$m$ & $m_1$ & $m_2$ & $H_m$\\
\hline
$2l+1$ for $1\le l\le k-1$ & 0 & $k-l$ & $(\infty, 0, -1, 1, -2, \ldots, -l)$\\
$2l$ for $2\le l\le k$ & 0 & $l-1$ & $(\infty, 0, -1, 1, -2, \ldots, l-1)$\\
$2k+1$ & 0 & 1 & $\Phi_0$\\
$2n-2l-1$ for $1\le l\le k-1$ & 0 & $k-l$ & $\Phi_0\cup\Phi_{k-l}\setminus (\infty, 0, -1, 1, -2, \ldots, -l)$\\
$2n-2l$ for $2\le l\le k$ & 0 & $l-1$ & $\Phi_0\cup\Phi_{l-1}\setminus (\infty, 0, -1, 1, -2, \ldots, l-1)$\\
\hline
\\
\multicolumn{4}{l}{$n=2k+2$, $V=\ZZ_{2k+1}\cup\{\infty\}$}\\ 
\hline
$m$ & $m_1$ & $m_2$ & $H_m$\\
\hline
$3$ & 0 & $1$ & $(0, -1, 1)$\\
$2l+1$ for $2\le l\le k$ & 0 & $k-l+1$ & $(\infty, 0, -1, 1, -2, \ldots, -l)$\\
$2l$ for $2\le l\le k$ & 0 & $l-1$ & $(\infty, 0, -1, 1, -2, \ldots, l-1)$\\
$2k+2$ & 0 & 1 & $\Phi_0$\\
$2n-3$ & 0 & $1$ & $\Phi_0\cup\Phi_1\setminus (0, -1, 1)$\\
$2n-2l-1$ for $2\le l\le k-1$ & 0 & $k-l+1$ & $\Phi_0\cup\Phi_{k-l}\setminus (\infty, 0, -1, 1, -2, \ldots, -l)$\\
$2n-2l$ for $2\le l\le k$ & 0 & $l-1$ & $\Phi_0\cup\Phi_{l-1}\setminus (\infty, 0, -1, 1, -2, \ldots, l-1)$\\
\hline
\end{tabular}
\end{table*}

Then, for $2n-3<m\le kn-3$, choose $1\le r\le k-2$ such that $3\le m-rn\le 2n-3$.
Let $m'=m-rn$ and choose $r$ Hamilton cycles $\Phi_{j_1},\Phi_{j_2},\ldots, \Phi_{j_{r}}$ such that $j_s\notin \{m'_1,m'_2\}$.
Then $H_{m'}\cup \left(\bigcup_{s=1}^r \Phi_{j_s}\right)$ is an eulerian graph of size $m$ since $H_m$ is even,
 contains a Hamilton cycle and is hence connected. 
\end{IEEEproof}

\vskip 5pt

\begin{IEEEproof}[Proof of Proposition \ref{girth4}]

The proposition can be readily verified for $n\in \{6,7\}$. 

First, we prove for the case $n$ even. 

Let $n'=n/2$ and $k=\floor{n'/2}$ and we show that there exists an eulerian graph of order $n$, girth at least four and size $m$,
for $n\le m\le nk$ and $m$ even, except for $m=nk-2$. The proof for $n$ even is similar to proof of Proposition \ref{evengraphs}. 

Consider the following collection of 
$k$ edge-disjoint Hamiliton cycles in $K_{n',n'}$ due to Dirac \cite{Dirac:1972}.

Let the vertex set $V=(\ZZ_{n'}\times \{\bullet,\circ\})$ and the partitions be $ \ZZ_{n'}\times \{\bullet\}$ and $\ZZ_{n'}\times\{\circ\})$.
Write $(a,b)$ as $a_b$ 
 and for $0\le i\le k-1$, consider the Hamiliton cycle $\Phi_i$ given by
\[ \Phi_i=(0_{\bullet},(2i)_{\circ},1_{\bullet},(1+2i)_\circ,\ldots, (n'-1)_\bullet,(n'-1+2i)_\circ).\]

As in Proposition \ref{evengraphs}, for $4\le m\le 2n-4$, there exist two Hamilton cycles $\Phi_{m_1}$ and $\Phi_{m_2}$ and a subgraph $H_m$ such that the following holds,
\begin{enumerate}[(i)]
\item $H_m$ is a subgraph of $\Phi_{m_1}\cup \Phi_{m_2}$,
\item $H_m$ is even with size $m$ and when $m\le n$, $H_m$ is connected and hence, eulerian.
\end{enumerate}
\noindent We give explicit constructions of $\Phi_{m_1}$, $\Phi_{m_2}$ and $H_m$ in Table \ref{tab:decomp2} and the rest of the proof proceeds in the same manner. Since the graphs constructed are subgraphs of $K_{n',n'}$, their girths are at least four.

\begin{table*}[htbp]
\centering
\renewcommand{\arraystretch}{1.3}
\setlength{\tabcolsep}{3pt}
\caption{Eulerian graphs of small size with order $n$, girth at least four}
\label{tab:decomp2}
\begin{tabular}{cccl}
\multicolumn{4}{l}{$n=4k$ or $n'=2k$, $V=\ZZ_{n'}\cup\{\bullet,\circ\}$}\\
\hline
$m$ & $m_1$ & $m_2$ & $H_m$\\
\hline
$4l$ for $1\le l\le k-1$ & 0 & $l$ & $(0_\bullet, 0_\circ, 1_\bullet, 1_\circ, \ldots, (2l-1)_\bullet, (2l-1)_\circ)$\\
$4l+2$ for $1\le l\le k-1$ & 0 & $l$ & $(0_\bullet, 0_\circ, 1_\bullet, 1_\circ, \ldots, (2l)_\bullet, (2l)_\circ)$\\
$4k$ & 0 & 1 & $\Phi_0$\\
$2n-4l$ for $1\le l\le k-1$ & 0 & $l$ & $\Phi_0\cup\Phi_{l}\setminus (0_\bullet, 0_\circ, 1_\bullet, 1_\circ, \ldots, (2l-1)_\bullet, (2l-1)_\circ)$\\
$2n-4l-2$ for $1\le l\le k-1$ & 0 & $l$ & $\Phi_0\cup\Phi_{l}\setminus (0_\bullet, 0_\circ, 1_\bullet, 1_\circ, \ldots, (2l)_\bullet, (2l)_\circ)$\\
\hline
\\
\multicolumn{4}{l}{$n=4k+2$ or $n'=2k+1$, $V=\ZZ_{n'}\cup\{\bullet,\circ\}$}\\ 
\hline
$m$ & $m_1$ & $m_2$ & $H_m$\\
\hline
$4l$ for $1\le l\le k-1$ & 0 & $l$ & $(0_\bullet, 0_\circ, 1_\bullet, 1_\circ, \ldots, (2l-1)_\bullet, (2l-1)_\circ)$\\
$4l+2$ for $1\le l\le k-1$ & 0 & $l$ & $(0_\bullet, 0_\circ, 1_\bullet, 1_\circ, \ldots, (2l)_\bullet, (2l)_\circ)$\\
$4k$ & 0 & 1 & $(0_\bullet, 2_\circ, 1_\bullet, 3_\circ, \ldots, (n'-2)_\bullet, 0_\circ)$\\
$4k+2$ & 0 & 1 & $\Phi_0$\\
$4k+4$ & 0 & 1 & $\Phi_0\cup\Phi_1\setminus (0_\bullet, 2_\circ, 1_\bullet, 3_\circ, \ldots, (n'-2)_\bullet, 0_\circ)$\\
$2n-4l$ for $1\le l\le k-1$ & 0 & $l$ & $\Phi_0\cup\Phi_{l}\setminus (0_\bullet, 0_\circ, 1_\bullet, 1_\circ, \ldots, (2l-1)_\bullet, (2l-1)_\circ)$\\
$2n-4l-2$ for $1\le l\le k-1$ & 0 & $l$ & $\Phi_0\cup\Phi_{l}\setminus (0_\bullet, 0_\circ, 1_\bullet, 1_\circ, \ldots, (2l)_\bullet, (2l)_\circ)$\\
\hline
\end{tabular}
\end{table*}

\vskip 3pt
Recall that $M(n)=2\floor{n^2/8}$ when $n$ is even. When $n=4k$, $M(n)=4k^2=nk$ and hence, the stated graphs are constructed.

When $n=4k+2$, note that $K_{2k,2k+2}$ (defined on partitions $\ZZ_{2k}\times \{\bullet\}$, $\ZZ_{2k+2}\times\{\circ\}$) is an eulerian graph with size $M(n)=4k^2+4k$ and girth at least four. Observe that $K_{2k,2k+2}$ contains cycles of even length $4\le m' \le 2k+2$, 
namely, $(0_\bullet, 0_\circ, 1_\bullet, 1_\circ, \ldots, (m'/2-1)_\bullet, (m'/2-1)_\circ)$. Hence, removing a cycle of length $m'$, we obtain eulerian graphs with order $n$ and girth at least four with size $m$, $nk-2 \le m \le M(n)-4$.

Finally, when $n$ is odd, let $m$ be odd, with $n\le m\le M(n)$ and $m\ne M(n)-2$. 
Then there exists an eulerian graph $H$ with order $n-1$, size $m-1$ and girth at least four.
Pick any edge in $H$ and subdivide the edge to obtain an eulerian graph with order $n$, size $m$ and girth at least four. 
This completes the proof.
\end{IEEEproof}

\vskip 5pt

\begin{figure}[hbtp]
\centering
\renewcommand{\arraystretch}{1.3}
\setlength{\tabcolsep}{10pt}

\begin{tabular}{cc}
 $H_{9,14}$ & $H_{9,16}$\\
 
\begin{tikzpicture}

\vertex[fill] (v0) at (10:1) {};
\vertex[fill] (v1) at (50:1) {};
\vertex[fill] (v2) at (90:1) {};
\vertex[fill] (v3) at (130:1) {};
\vertex[fill] (v4) at (170:1) {};
\vertex[fill] (v5) at (210:1) {};
\vertex[fill] (v6) at (250:1) {};
\vertex[fill] (v7) at (290:1) {};
\vertex[fill] (v8) at (330:1) {};

\Edge[](v0)(v1)
\Edge[](v0)(v5)
\Edge[](v0)(v7)
\Edge[](v0)(v8)
\Edge[](v1)(v2)
\Edge[](v2)(v3)
\Edge[](v2)(v5)
\Edge[](v2)(v7)
\Edge[](v3)(v4)
\Edge[](v4)(v5)
\Edge[](v4)(v7)
\Edge[](v4)(v8)
\Edge[](v5)(v6)
\Edge[](v6)(v7)

\end{tikzpicture}

&

\begin{tikzpicture}

\vertex[fill] (v0) at (10:1) {};
\vertex[fill] (v1) at (50:1) {};
\vertex[fill] (v2) at (90:1) {};
\vertex[fill] (v3) at (130:1) {};
\vertex[fill] (v4) at (170:1) {};
\vertex[fill] (v5) at (210:1) {};
\vertex[fill] (v6) at (250:1) {};
\vertex[fill] (v7) at (290:1) {};
\vertex[fill] (v8) at (330:1) {};

\Edge[](v0)(v1)
\Edge[](v0)(v5)
\Edge[](v0)(v7)
\Edge[](v0)(v8)
\Edge[](v1)(v2)
\Edge[](v1)(v4)
\Edge[](v1)(v6)
\Edge[](v2)(v3)
\Edge[](v2)(v5)
\Edge[](v2)(v8)
\Edge[](v3)(v4)
\Edge[](v4)(v5)
\Edge[](v4)(v8)
\Edge[](v5)(v6)
\Edge[](v6)(v7)
\Edge[](v6)(v8)
\end{tikzpicture}
\end{tabular}
\caption{Eulerian Graphs of order $9$ and size $14,16$ }
\label{fig:smallgraphs}
\end{figure}
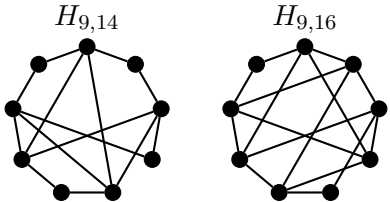

\begin{IEEEproof}[Proof of Proposition \ref{girth4opp}]
Eulerian graphs with order nine, girth four and sizes 14, 16 are given in Figure \ref{fig:smallgraphs}.
For each graph of order nine, subdivide any edge to obtain an eulerian graph of order ten, girth four and orders 15, 17.  
Denote these graphs by $H_{n,m}$, where $n$ is the order and $m$ is the size. 

For $n\ge 11$, let $n'=2\lfloor (n-1)/2\rfloor$. 
%Consider the complete bipartite graph $G=K_{2\lfloor n'/4\rfloor,2\lceil n'/4\rceil}$. Then $G$ is a graph of order $n'$, girth four and size $M
Then $K_{2\lfloor n'/4\rfloor,2\lceil n'/4\rceil}$ is a graph of order $n'$, girth four and size $M(n')$, 
containing a subgraph $K_{4,4}$. Replacing the subgraph $K_{4,4}$ with 
\begin{equation*}
\begin{cases}
H_{9,14}\mbox{ or }H_{9,16},		& \mbox{if $n$ is odd,}\\
H_{10,15}\mbox{ or }H_{10,17},	& \mbox{otherwise,}
\end{cases}
\end{equation*}
\noindent yields an eulerian graph of order $n$, girth at least four with the desired sizes.
\end{IEEEproof}

\section{MDS Symbol-Pair Codes via Development}\label{sec:develop}

%\begin{prop} \label{prop:g2dev}
%Let $n$ be even.
%If a $\Gamma^2$-development $(n,n-1)$-symbol-pair code exists with $|\Gamma|=q$, then
%an MDS $(n,n-1)_q$-symbol-pair code exists.
%\end{prop}

We provide detailed proofs of propositions given in Subsection \ref{sec:nonlinear}.

\begin{IEEEproof}[Proof of Proposition \ref{prop:g2dev}]

It is readily verified that $|\C|=q^{3}$ and so, it remains to show that $\C$ has minimum pair-distance $n-1$.

Suppose otherwise that there exist distinct codewords $\phi(\vu,\alpha,\alpha')$ and $\phi(\vv,\beta,\beta')$ in $\C$ 
with 
\begin{equation*}
D_\text{p}(\phi(\vu,\alpha,\alpha'),\phi(\vv,\beta,\beta'))<n-1.
\end{equation*} 
Then there exist 
$i,j\in \ZZ_n$, $i\ne j$, such that
{\small
\begin{align*}
(\phi(\vu,\alpha,\alpha')_{i},\phi(\vu,\alpha,\alpha')_{i+1}) &=(\phi(\vv,\beta,\beta')_{i},\phi(\vv,\beta,\beta')_{i+1}),\\
(\phi(\vu,\alpha,\alpha')_{j},\phi(\vu,\alpha,\alpha')_{j+1}) &=(\phi(\vv,\beta,\beta')_{j},\phi(\vv,\beta,\beta')_{j+1}).
\end{align*}
}
Without loss of generality, assume $i\equiv 0\bmod 2$. Suppose $j\equiv 0\bmod 2$. Then 
\begin{align*}
(u_i+\alpha,u_{i+1}+\alpha') &= (v_i+\beta,v_{i+1}+\beta') ,\\
(u_j+\alpha,u_{j+1}+\alpha') &= (v_j+\beta,v_{j+1}+\beta').
\end{align*}
Hence, 
\begin{equation*}
(u_i-u_j,u_{i+1}-u_{j+1})= (v_i-v_j,v_{i+1}-v_{j+1}),%, &\mbox{ or,}\\
\end{equation*}
\noindent contradicting Condition (i) in Definition \ref{def:g2dev}.

Similarly, when  $j \equiv 1 \bmod 2$, 
\begin{align*}
(u_i+\alpha,u_{i+1}+\alpha') &= (v_i+\beta,v_{i+1}+\beta') ,\\
(u_j+\alpha',u_{j+1}+\alpha) &= (v_j+\beta',v_{j+1}+\beta),
\end{align*}
\noindent and so,
\begin{equation*}
(u_i-u_{j+1},u_{i+1}-u_{j})= (v_i-v_{j+1},v_{i+1}-v_{j}).%, &\mbox{ or,}\\
\end{equation*}
\noindent We derive a contradiction to 
Condition (ii) in Definition \ref{def:g2dev}.
\end{IEEEproof}

\vskip 5pt

%\begin{rem}~~
%\begin{enumerate}
%\item As remarked earlier, the method of developing is similar to the use of generator matrices. 
%Indeed, suppose $\Gamma=\ZZ_q$ and $\C_0$ is a $\ZZ_q$-developing $(n,n-1)$-symbol-pair code with $\vc\in \C_0$, 
%such that 
%\begin{equation*}
%\C_0=\{\alpha \vc: \alpha\in \ZZ_q\}.
%\end{equation*}
%Then 
%\begin{equation*} 
%\C=\{\phi(\vu,\alpha,\alpha'):\vu \in\C_0,\alpha,\alpha' \in \ZZ_q, \phi \mbox{ defined in (\ref{eq:phi})}\}
%\end{equation*}
%is in fact $\ZZ_q$-linear with generator matrix,
%\begin{equation*}
%\left( 
%\begin{array}{ccccccc}
%1 & 0 & 1 & 0 & \cdots & 1 & 0\\
%0 & 1 & 0 & 1 & \cdots & 0 & 1\\
%c_0 & c_1 & c_2 & c_3 & \cdots & c_{n-2} & c_{n-1}
%\end{array} 
%\right).
%\end{equation*}
%
%\item On the other hand, the method of developing, in some sense, is necessary to construct MDS codes for certain parameters. 
%For example, 
%Corollary \ref{cor:2p} exhibits the existence of MDS $(8,7)_{2p}$-symbol-pair code via the method of developing,  
%but Corollary \ref{cor:nonexist} demonstrates the nonexistence of $\ZZ_{2p}$-linear MDS $(8,7)_{2p}$-symbol-pair code. 
%%hence, the method of developing is in some sense necessary.
%\end{enumerate}
%\end{rem}

%\vskip 5pt
%\begin{prop}\label{prop:2p}
%Let $p$ be prime with $p\ge 5$. Then a $(\ZZ_p\times \ZZ_2)^2$-development $(8,7)$-symbol-pair code exists.
%\end{prop}

\begin{IEEEproof}[Proof of Proposition \ref{prop:2p}]
%Let $\C_0$ consists of the following four codewords,
%\begin{align*}
%((0,0),(0,0), (0,0),(1,0), (0,0),(1,1), (0,0),(0,1)),\\
%((0,0),(0,0), (0,1),(1,1), (2,0),(0,1), (2,1),(2,0)),\\
%((0,0),(0,0), (1,0),(0,0), (1,1),(0,0), (0,1),(0,0)),\\
%((0,0),(0,0), (1,1),(0,1), (0,1),(2,0), (2,0),(2,1)).\\
%\end{align*}
%Let $\C_1$ be the following set of $2p-4$ codewords,
%\begin{align*}
%((0,0),(0,0), (a,0),(\hat a,1), 	(3a,1),(0,1), (2a,1),(2\hat a,0)),\\
%((0,0),(0,0), (a,1),(a,0), 		(0,1),(3a,1), (2a,0),(2a,1)),\\
%\end{align*}
%\noindent where $a\in\{2,3,\ldots, p-1\}$ and 
%\begin{equation*}
%\hat a=\begin{cases}
%p-1,	&\mbox{if $a=2$,}\\
%a-1, &\mbox{otherwise.}
%\end{cases}
%\end{equation*}

We exhibit that $\C$ is a $(\ZZ_p\times \ZZ_2)^2$-development $(8,7)$-symbol-pair code,
by checking the conditions of Definition \ref{def:g2dev}. 

The values of $u_i-u_{i+2}$ for $\vu\in\C$, $i\in\ZZ_8$ are given in Table \ref{tab:condi} and we verify that for $i\in\ZZ_8$
\begin{equation}\label{eq:diff}
u_i-u_{i+2}\ne v_i-v_{i+2} \mbox{ for $\vu,\vv\in \C$.}
\end{equation}

For Condition (i), note that when $j=i+2$, (\ref{eq:diff}) ensures that the differences $(u_i-u_{i+2},u_{i+1}-u_{i+3})$ are distinct.
Hence, it remains to check when $i-j\equiv 4\bmod 8$ and these values are given in Table \ref{tab:condii}.
%when $(i,j)\in \{(0,4),(1,5),(2,6),(3,7)\}$ and these values are given in Table \ref{tab:condii}.

For Condition (ii), if $i\not\equiv j \bmod 2$, then 
either $j+1=i+2$, $i+1=j+2$, $j=i+3$ or $i=j+3$ since $n=8$. 
(\ref{eq:diff}) ensures that the values $(u_i-u_{j+1},u_{i+1}-u_{j})$ are distinct.
%Suppose $j+1=i+2$, then 
%
%The remaining cases can be argued similarly.
\end{IEEEproof}

\begin{table*}[hbtp]
\centering
\caption{Differences $u_i-u_{i+2}$ for $\vu\in\C$, $i\in\ZZ_8$}
\label{tab:condi}
\renewcommand{\arraystretch}{1.1}
\setlength{\tabcolsep}{5pt}
\begin{tabular}{cll}
\hline
& \multicolumn{2}{c}{$u_i-u_{i+2}$}\\
$i$& $\C_0$ & $\C_1$\\
\hline
0 & $\{ (0,0), (0,1), (-1,0), (-1,1) \}$ & 	 $\{ (-a,0), (-a,1) \}$ for $a\in \{2,3,\ldots,p-1\}$\\
1 & $\{ (-1,0), (-1,1), (0,0), (0,1) \}$ & 	 $\{ (-\hat a,1), (-a,0) \}$ for $a\in \{2,3,\ldots,p-1\}$\\
2 & $\{ (0,0), (-2,1), (0,1), (1,0) \}$ & 	 $\{ (-2a,1), (a,0) \}$ for $a\in \{2,3,\ldots,p-1\}$\\
3 & $\{ (0,1), (1,0), (0,0), (-2,1) \}$ & 	 $\{ (\hat a,0), (-2a,1) \}$ for $a\in \{2,3,\ldots,p-1\}$\\
4 & $\{ (0,0), (0,1), (1,0), (-2,1) \}$ & 	 $\{ (a,0), (-2a,1) \}$ for $a\in \{2,3,\ldots,p-1\}$\\
5 & $\{ (1,0), (-2,1), (0,0), (0,1) \}$ & 	 $\{ (-2\hat a,1), (a,0) \}$ for $a\in \{2,3,\ldots,p-1\}$\\
6 & $\{ (0,0), (2,1), (0,1), (2,0) \}$ & 	 $\{ (2a,1), (2a,0) \}$ for $a\in \{2,3,\ldots,p-1\}$\\
7 & $\{ (0,1), (2,0), (0,0), (2,1) \}$ & 	 $\{ (2\hat a,0), (2a,1) \}$ for $a\in \{2,3,\ldots,p-1\}$\\
\hline
\end{tabular}
\end{table*}

\begin{table*}[hbtp]
\centering
\scriptsize
\caption{Differences $(u_i-u_{j},u_{i+1}-u_{j+1})$ for $\vu\in\C$ and $i-j \equiv 4 \bmod 8$}
\label{tab:condii}
\renewcommand{\arraystretch}{1.1}
\setlength{\tabcolsep}{5pt}
\begin{tabular}{cll}
\hline
& \multicolumn{2}{c}{ $(u_i-u_{j},u_{i+1}-u_{j+1})$}\\
$(i,j)$& $\C_0$ & $\C_1$\\
\hline
(0,4)& $\{ ((0,0),(-1,1)), ((-2,0),(0,1)), ((-1,1),(0,0)), ((0,1),(-2,0)) \}$ & 	 $\{ ((-3a,1),(0,1)), ((0,1),(-3a,1)) \}$ for $a\in \{2,3,\ldots,p-1\}$\\
(1,5)& $\{ ((-1,1),(0,0)), ((0,1),(-2,0)), ((0,0),(1,1)), ((-2,0),(-1,1)) \}$ & 	 $\{ ((0,1),(-a,1)), ((-3a,1),(-a,1)) \}$ for $a\in \{2,3,\ldots,p-1\}$\\
(2,6)& $\{ ((0,0),(1,1)), ((-2,0),(-1,1)), ((1,1),(0,0)), ((-1,1),(-2,0)) \}$ & 	 $\{ ((-a,1),(-\hat a,1)), ((-a,1),(-a,1)) \}$ for $a\in \{2,3,\ldots,p-1\}$\\
(3,7)& $\{ ((1,1),(0,0)), ((-1,1),(2,0)), ((0,0),(1,1)), ((-2,0),(0,1)) \}$ & 	 $\{ ((-\hat a,1),(3a,1)), ((-a,1),(0,1)) \}$ for $a\in \{2,3,\ldots,p-1\}$\\
\hline
\end{tabular}
\end{table*}

\section*{Acknowledgement}

We are grateful to the anonymous reviewers and Associate Editor Professor Kashyap 
for their constructive comments,
which improved greatly the presentation of this paper.

\bibliographystyle{IEEEtran}

\begin{thebibliography}{10}
\providecommand{\url}[1]{#1}
\csname url@samestyle\endcsname
\providecommand{\newblock}{\relax}
\providecommand{\bibinfo}[2]{#2}
\providecommand{\BIBentrySTDinterwordspacing}{\spaceskip=0pt\relax}
\providecommand{\BIBentryALTinterwordstretchfactor}{4}
\providecommand{\BIBentryALTinterwordspacing}{\spaceskip=\fontdimen2\font plus
\BIBentryALTinterwordstretchfactor\fontdimen3\font minus
  \fontdimen4\font\relax}
\providecommand{\BIBforeignlanguage}[2]{{%
\expandafter\ifx\csname l@#1\endcsname\relax
\typeout{** WARNING: IEEEtran.bst: No hyphenation pattern has been}%
\typeout{** loaded for the language `#1'. Using the pattern for}%
\typeout{** the default language instead.}%
\else
\language=\csname l@#1\endcsname
\fi
#2}}
\providecommand{\BIBdecl}{\relax}
\BIBdecl

\bibitem{Cheeetal:2012}
Y.~M. Chee, H.~Kiah, and C.~Wang, ``Maximum distance separable symbol-pair codes,'' in 
\emph{ISIT 2012 --  Proceedings of the 2012 IEEE International Symposium on Information Theory},
  \hskip 1em plus 0.5em minus 0.4em\relax Cambridge, Massachusetts: IEEE Press, 2012, pp. 2896--2900.

\bibitem{CassutoBlaum:2010}
Y.~Cassuto and M.~Blaum, ``Codes for symbol-pair read channels,'' in \emph{ISIT
  2010 -- Proceedings of the 2010 IEEE International Symposium on Information
  Theory}.\hskip 1em plus 0.5em minus 0.4em\relax Austin, Texas: IEEE Press,
  June 2010, pp. 988--992.

\bibitem{CassutoBlaum:2011}
------, ``Codes for symbol-pair read channels,'' \emph{IEEE Trans. Inform.
  Theory}, vol.~57, no.~12, pp. 8011--8020, 2011.

\bibitem{CassutoLitsyn:2011}
Y.~Cassuto and S.~Litsyn, ``Symbol-pair codes: algebraic constructions and
  asymptotic bounds,'' in \emph{ISIT 2011 -- Proceedings of the 2011 IEEE
  International Symposium on Information Theory}.\hskip 1em plus 0.5em minus
  0.4em\relax St Petersburg, Russia: IEEE Press, July 2011, pp. 2348--2352.

  \bibitem{Yaakobietal:2012}
E.~Yaakobi and J.~Bruck and P.H.~Siegel, ``Decoding of cyclic codes over symbol-pair read channels,''
\emph{ISIT 2012 --  Proceedings of the 2012 IEEE International Symposium on Information Theory},
\hskip 1em plus 0.5em minus 0.4em\relax Cambridge, Massachusetts: IEEE Press, 2012, pp. 2891--2895.



  



\bibitem{Hedayatetal:1999}
A.~S. Hedayat, N.~J.~A. Sloane, and J.~Stufken, \emph{Orthogonal Arrays}, ser.
  Springer Series in Statistics.\hskip 1em plus 0.5em minus 0.4em\relax New
  York: Springer-Verlag, 1999.




\bibitem{BondyMurty:2008}
J.~A. Bondy and U.~S.~R. Murty, \emph{Graph Theory}, ser. Graduate Texts in
  Mathematics.\hskip 1em plus 0.5em minus 0.4em\relax Springer, 2008.




\bibitem{Novak:1971}
J.~Nov{\'a}k, ``Eulerovsk{\'e} grafy bez troj{\'u}heln{\'\i}k{\r{u}} s
  maxim{\'a}ln{\'\i}m po{\v{c}}tem hran,'' \emph{Sborn{\'\i}k
  v{\v{e}}deck{\'y}ch prac{\'\i} {V{\v{S}}ST}, Liberec}, 1971.

\bibitem{Novak:1974}
------, ``Edge bases of complete uniform hypergraphs,'' \emph{Mat. \v Casopis
  Sloven. Akad. Vied}, vol.~24, pp. 43--57, 1974.

  \bibitem{Bethetal:1999}
T.~Beth, D.~Jungnickel, and H.~Lenz, \emph{Design Theory}, 2nd~ed.\hskip 1em
  plus 0.5em minus 0.4em\relax Cambridge University Press, 1999.


\bibitem{Dirac:1972}
\BIBentryALTinterwordspacing
G.~A. Dirac, ``On hamilton circuits and hamilton paths,'' \emph{Mathematische
  Annalen}, vol. 197, pp. 57--70, 1972.
  



\end{thebibliography}

\begin{IEEEbiographynophoto}
{\bf Yeow Meng Chee}(SMÕ08) received the B.Math. degree in computer science and combinatorics and optimization 
and the M.Math. and Ph.D. degrees in computer science, 
from the University of Waterloo, Waterloo, ON, Canada, in 1988, 1989, and 1996, respectively.

Currently, he is an Associate Professor at the Division of Mathematical Sciences, School of Physical and Mathematical Sciences, 
Nanyang Technological University, Singapore. 
Prior to this, he was Program Director of Interactive Digital Media R\&D in the Media Development Authority of Singapore, 
Postdoctoral Fellow at the University of Waterloo and IBM's Z\"urich Research Laboratory, 
General Manager of the Singapore Computer Emergency Response Team, and 
Deputy Director of Strategic Programs at the Infocomm Development Authority, Singapore. 
His research interest lies in the interplay between combinatorics and computer science/engineering, 
particularly combinatorial design theory, coding theory, extremal set systems, and electronic design automation.
\end{IEEEbiographynophoto}

\begin{IEEEbiographynophoto}
{\bf Lijun Ji} received the Ph.D. degree in applied mathematics from Suzhou University, China, in 2003. 
He is currently a Professor at Department of Mathematics in Suzhou University. 
His research interests include combinatorial design theory and coding theory.
\end{IEEEbiographynophoto}

\begin{IEEEbiographynophoto}
{\bf Han Mao Kiah} received the B.Sc.(Hon) degree in mathematics from the National University of Singapore, Singapore in 2006. 
Currently, he is working towards his Ph.D. degree at the Division of Mathematical Sciences, School of Physical and Mathematical Sciences, Nanyang Technological University, Singapore.
His research interest lies in the application of combinatorics to engineering problems in information theory. 
In particular, his interests include combinatorial design theory, coding theory and power line communications.
\end{IEEEbiographynophoto}

\begin{IEEEbiographynophoto}
{\bf Chengmin Wang} received the B.Math. and Ph.D. degrees in mathematics from Suzhou University, China in 2002 and 2007, respectively.
Currently, he is an Associate Professor at the School of Science, Jiangnan University, China. 
Prior to this, he was a Visiting Scholar at the School of Computing, Informatics and Decision Systems Engineering, 
Arizona State University, USA, from 2010 to 2011 and was a 
Research Fellow at the Division of Mathematical Sciences, 
School of Physical and Mathematical Sciences, Nanyang Technological University, Singapore from 2011 to 2012. 
His research interests include combinatorial design theory and its applications in coding theory and cryptography.
\end{IEEEbiographynophoto}

\begin{IEEEbiographynophoto}
{\bf Jianxin Yin} graduated from Suzhou University, Suzhou, China, in 1977.
Since 1977, he has been a Teacher in the Department of Mathematics, Suzhou University.  
Currently, he is a Full Professor there. He is an editorial board member of Journal of Combinatorial Designs 
and an associate editor of Discrete Mathematics, Algorithms and Applications. 
He has held various grants of the National Natural Science Foundation of China (NSFC) as Project Leader. 
His research interests include combinatorial design theory, combinatorial coding theory
and the application of  combinatorics to software testing..
\end{IEEEbiographynophoto}

\end{document}